\documentclass{article}
\usepackage{Main,amsmath,graphicx}
\pdfoutput=1

\usepackage[ruled,vlined]{algorithm2e}
\usepackage{graphicx}
\usepackage{amssymb}
\usepackage{url}
\usepackage{color}
\usepackage{amsmath}
\usepackage{stmaryrd}
\usepackage{tikz}
\usepackage{listings}
\usepackage{amsthm}
\usepackage{hyperref}
\usepackage{url}
\usepackage{booktabs}
\usepackage{multirow}
\usepackage{siunitx}
\usepackage{makecell}
\usepackage{wrapfig}

\usepackage[utf8]{inputenc}

\title{TENSOR SHAPE SEARCH FOR OPTIMUM DATA COMPRESSION}

\name{Ryan Solgi$^{\star\dagger}$  \qquad Zichang He$^{\star}$ \qquad William Jiahua Liang$^{\ddag}$ \qquad Zheng Zhang$^{\star}$\thanks{This work is supported by NSF Grants CCF-1817037 and CCF-1763699, and the Department of Geography, University of California Santa Barbara.}}
\address{$^{\star}$ Department of Electrical and Computer Engineering, UC Santa Barbara\\$^{\dagger}$ Department of Geography, UC Santa Barbara\\
$^{\ddag}$ School of Engineering \& Applied Science, University of Pennsylvania
}
\begin{document}

\ninept
\maketitle
\begin{abstract}
Various tensor decomposition methods have been proposed for data compression. In real world applications of the tensor decomposition, selecting the tensor shape for the given data poses a challenge and the shape of the tensor may affect the error and the compression ratio. In this work, we study the effect of the tensor shape on the tensor decomposition and propose an optimization model to find an optimum shape for the tensor train (TT) decomposition. The proposed optimization model maximizes the compression ratio of the TT decomposition given an error bound. We implement a genetic algorithm (GA) linked with the TT-SVD algorithm to solve the optimization model. We apply the proposed method for the compression of RGB images. The results demonstrate the effectiveness of the proposed evolutionary tensor shape search for the TT decomposition.
\end{abstract}
\begin{keywords}
Tensor decomposition, tensor train decomposition, data compression, genetic algorithm, evolutionary algorithm
\end{keywords}

\section{Introduction}
Compressing high-volume data via tensor decomposition has gained great success in the recent years. The tensor decomposition is applied to approximate high dimensional problems in different domains including scientific computation, machine learning, and visual processing \cite{Kolda, Zhang_a,Zhang_b, Zhang_c, zhang2021fpga}. In order to decompose a high order tensor into the low-dimensional parameters, different methods have been successfully applied including the CANDECOMP/PARAFAC (CP) decomposition \cite{Bro}, the Tucker decomposition \cite{Tucker}, and the tensor train (TT) decomposition \cite{Oseledets}. The tensor decomposition was also extended to a more general form called the tensor networks which leads to various decomposition formats \cite{Li}. In recent years, Bayesian methods have also been developed for automatic rank determination in various tensor problems including tensor completion and tensorized neural network training \cite{Zhao_a, hawkins2020towards, Hawkins}.

Regardless of the specific choice of a tensor decomposition method, the data or the model parameters are represented as a $d$-way tensor prior to the decomposition. This often involves a reshaping step which changes the dimension and mode size of a tensor without changing the total number of its elements. The shape of a tensor affects the accuracy and the compression ratio of the subsequent tensor decomposition. Despite the importance of finding an optimum shape for the tensor decomposition, studies on this domain remains very sparse.

In this study we investigate the effect of the tensor shape on the compression ratio achieved by the tensor decomposition. We formulate the task of finding the best shape for the tensor decomposition as an optimization model. We present a genetic algorithm (GA) for solving the problem. Specifically, we narrow down the study to the TT decomposition, but our proposed technique can be extended to other tensor decomposition methods.
\section{Related works}
\subsection{Tensor Decomposition and Applications} A detailed review of tensor decomposition and its application in different fields (e.g.,  signal processing, computer vision, data mining, scientific computing and neuroscience) is provided in \cite{Kolda}. Furthermore, the tensor decomposition has been recently shown promising in various areas including neural network (NN) compression \cite{Kim, Lebedev, Nikov, Wang_W, Shupeng, Jingling, Hawkins}, uncertainty quantification \cite{He, Zhang_a, Zhang_b,Zhang_c}, and tensor completion/recovery \cite{Rai, Zhao_a, Zhao_b} to name a few. In many real-world applications, deciding about some hyper-parameters (such as tensor ranks and tensor shapes) of the tensor decomposition can be very challenging. There have been some recent studies which addressed the tensor rank determination problem \cite{Zhao_a, Zhao_b, Hawkins, He,hawkins2020towards}. The recent work~\cite{hawkins2020towards} determines the tensor ranks automatically in a neural network training, enabling on-device training of neural networks with limited computing resources~\cite{zhang2021fpga}. However, the study of the effect of the shape on the tensor decomposition has been rarely reported in the literature. Therefore, in this work we investigate how reshaping may affect the result of the decomposition and how an optimum shape can be found if there exists one.

\subsection{Evolutionary Algorithms}The origin of evolutionary computation dates back to mid 1950s when it was applied in mathematical programming, machine learning, and industrial manufacturing and notably the invention of evolutionary strategies (ES), evolutionary programming (EP) and genetic algorithms (GAs) \cite{solgi_a}. \cite{holland_ga} presented the early version of the genetic algorithm (GA). Over the past years, variations of evolutionary algorithms (i.e., GAs) have been developed and have been extensively applied to solve problems in various fields where the problems were not approachable with other optimization methods. \cite{Bozorg-Haddad} presented a wide range of evolutionary algorithms including GAs and their applications in engineering domains. Particularly, \cite{Huang} applied an evolutionary algorithm to find optimal hyper parameters of the singular value decomposition for the neural network compression.

\subsection{Evolutionary Algorithms and Tensor Decomposition} At the intersection of evolutionary algorithms and tensor decompositions, \cite{Marzullo} suggested tensor decomposition-based mutation in the neuroevolution of augmenting topoplogies (NEAT) algorithm. \cite{Wang_Q} applied the CP decomposition to solve high-dimensional optimization problems with evolutionary algorithms. \cite{laura} formulated the CP decomposition of non-negative tensors as a stochastic problem and solved it using an evolutionary algorithm. Also, \cite{Li} applied an evolutionary search for determining an optimum tensor network topology. To the best of our knowledge, the present study is the first endeavor which applies an evolutionary tensor shape search with the tensor decomposition for an optimum data compression.

\section{Background}
Throughout this manuscript, capital calligraphic letters (e.g., {$\mathcal{A}$}) are used to denote tensors, boldface capital letters (e.g., $\textbf{A}$) are used for matrices, boldface lower case letters (e.g., $\boldsymbol{a}$) are used for vectors, and Roman (e.g., $a$) or Greek (e.g., $\alpha$) letters are used for scalars.  $\mathcal{A}[i_1,...,i_d]$ refers to the element $i_1,...,i_d$ of the tensor $\mathcal{A}$.

\subsection{Tensor Shape and Reshaping} An order-$k$ ($k$-way) tensor \textbf{$\mathcal{X}$} $\in$ $\mathbb{R}^{I_1\times ...\times I_k}$ denotes a $k$-dimensional data array. The order of a tensor is the number of its dimensions. We use $\boldsymbol\theta=(I_1, I_2, \cdots, I_k)$ to specify the shape of a tensor, where $I_j\in \mathbb{N}$ is the size of dimension $j$. Therefore, the shape of a tensor determines the order and the number of elements on each dimension. Reshaping refers to changing the order and the number of elements on each dimension. For example, a $k$-way tensor \textbf{$\mathcal{X}$} $\in$ $\mathbb{R}^{I_1\times ...\times I_k}$ can be reshaped to a $d$-way tensor like \textbf{$\mathcal{Y}$} $\in$ $\mathbb{R}^{n_1\times ...\times n_d}$. Reshaping a tensor may change its cardinality. If the cardinality of \textbf{$\mathcal{Y}$}, $\vert\textbf{$\mathcal{Y}$}\vert$, is greater than that of \textbf{$\mathcal{X}$} ($\vert\textbf{$\mathcal{Y}$}\vert>\vert\textbf{$\mathcal{X}$}\vert$), then dummy elements (e.g., zeros) are entered. Throughout the manuscript, we applied two different functions for reshaping: (1) ${\rm reshape}(\textbf{$\mathcal{X}$}, \boldsymbol\theta)$ is used when reshaping does not change the cardinality, and (2) $\Phi(\textbf{$\mathcal{X}$},\boldsymbol\theta)$ is used to denote reshaping a given tensor \textbf{$\mathcal{X}$} to a new shape $\boldsymbol\theta$ if reshaping can change the cardinality.


\subsection{Tensor Train (TT) Decomposition}
In the tensor train (TT) format \cite{Oseledets}, a $d$-way tensor $\textbf{$\mathcal{Y}$} \in \mathbb{R}^{n_1\times .... \times n_d}$ is approximated with a set of $d$ cores $\bar{\textbf{$\mathcal{G}$}}=\{\textbf{$\mathcal{G}$}_1, \textbf{$\mathcal{G}$}_2, ..., \textbf{$\mathcal{G}$}_d\}$ where $\textbf{$\mathcal{G}$}_j \in \mathbb{R}^{r_{j-1}\times n_j \times r_{j}}$ , $r_j$'s for $j=1,...,d-1$ are the ranks, $r_0=r_d=1$, and each element of $\textbf{$\mathcal{Y}$}$ is approximated by Eq.~\eqref{eq:TT_format}:
\begin{align} 
\label{eq:TT_format}
\hat{\textbf{{$\mathcal{Y}$}}}[i_1,...,i_d]=\sum_{l_0,...,l_d} {\textbf{$\mathcal{G}$}_1[l_0,i_1,l_1]\textbf{$\mathcal{G}$}_2[l_1,i_2,l_2]...\textbf{$\mathcal{G}$}_d[l_{d-1},i_d,l_d]}. \end{align}

Given an error bound ($\epsilon$), the core factors, \textbf{$\mathcal{G}_j$}'s, are computed using ($d-1$) sequential singular value decomposition (SVD) of the auxiliary matrices formed by unfolding tensor \textbf{$\mathcal{Y}$} along different axes. This decomposition process which is called the TT-SVD is presented in Algorithm \ref{TT-SVD}.

\begin{algorithm}[t]
  \caption{TT-SVD}
  \label{TT-SVD}
  \KwIn{$d$-way tensor \textbf{$\mathcal{Y}$}, error bound $\epsilon$.}
  \KwOut{$\bar{\textbf{$\mathcal{G}$}}=\{\textbf{$\mathcal{G}$}_1, \textbf{$\mathcal{G}$}_2, ..., \textbf{$\mathcal{G}$}_d\}$}
  $\sigma=\frac{\epsilon}{d-1}\Vert \textbf{$\mathcal{Y}$} \Vert_F$\\
  $r_0=1$, 
  $r_d=1$,  
 $\textbf{W}=$ reshape$(\mathcal{Y},(n_1,\frac{\vert \mathcal{Y} \vert}{n_1}))$\\
 \For{$j=1$ to $j=d-1$}{
 $\textbf{W}=$ reshape$(\textbf{W},(r_{j-1}n_j,\frac{\vert \textbf{W} \vert}{r_{j-1}n_j}))$\\
 Compute $\sigma$-truncated SVD: $\textbf{W}=\textbf{USV}^T+\textbf{E}$, where $\Vert \textbf{E} \Vert_F \leq \sigma$\\
 $r_j$ = the rank of matrix \textbf{W} based on $\sigma$-truncated SVD\\
 $\textbf{$\mathcal{G}$}_j=$ reshape$(\textbf{U},(r_{j-1},n_j,r_j))$\\
 $\textbf{W}=\textbf{SV}^T$\\
 }
 $\textbf{$\mathcal{G}$}_d=$ reshape$(\textbf{W},(r_{d-1},n_d,r_d))$\\
 Return $\bar{\textbf{$\mathcal{G}$}}=\{\textbf{$\mathcal{G}$}_1, \textbf{$\mathcal{G}$}_2, ..., \textbf{$\mathcal{G}$}_d\}$\\
\end{algorithm}

In this work, we only apply the proposed evolutionary tensor shape search for the TT-SVD. However, it is possible to extend this framework to other tensor decomposition methods such as the CP decomposition, the Tucker decomposition, and generally to the tensor networks. 

\section{Tensor shape optimization}
Let \textbf{$\mathcal{X}$} $\in$ $\mathbb{R}^{I_1\times ...\times I_k}$ be the original data given to be compressed using the TT decomposition and $\hat{\textbf{$\mathcal{X}$}} \in \mathbb{R}^{I_1\times ...\times I_k}$ is the approximation of the given \textbf{$\mathcal{X}$}. For example, \textbf{$\mathcal{X}$} can be an RGB image where $k=3$. In order to compress the given data, the first step is reshaping the given \textbf{$\mathcal{X}$} into a $d$-way tensor (usually $d \geq k$) like \textbf{$\mathcal{Y}$} $\in$ $\mathbb{R}^{n_1\times n_2\times ...\times n_d}$. If the cardinality of \textbf{$\mathcal{Y}$}, $\vert\textbf{$\mathcal{Y}$}\vert$, is greater than that of \textbf{$\mathcal{X}$} ($\vert\textbf{$\mathcal{Y}$}\vert>\vert\textbf{$\mathcal{X}$}\vert$), then dummy elements (e.g., zeros) are entered. Next, \textbf{$\mathcal{Y}$} is approximated using the TT decomposition where $\hat{\textbf{$\mathcal{Y}$}}$ $\in$ $\mathbb{R}^{n_1\times n_2\times ...\times n_d}$ is the approximation of \textbf{$\mathcal{Y}$}. 

Given $d$ (the order of \textbf{$\mathcal{Y}$}), let us define $\boldsymbol{\theta}=(n_1,n_2,\cdots,n_d)$ as a possible shape; $\textbf{$\mathcal{Y}$}_{\boldsymbol{\theta}}$ and $\hat{\textbf{$\mathcal{Y}$}}_{\boldsymbol{\theta}}$ refer to the tensor with shape  $\boldsymbol{\theta}$ and its approximation, respectively; and let $S$ be the space made of all possible $\boldsymbol{\theta}$'s such that  $n_i \in \mathbb{N}$ and $l \leq n_i \leq u$ for $i=1,2,..,d$ and $l, u\in \mathbb{N}$. If $l=1$, $d$ is the maximum order because when $n_i=1$ dimension $i$ becomes ineffective, practically.
We are looking for a tensor shape which maximizes the compression ratio of the TT decomposition given an error bound $\epsilon$ as below: 
\begin{align}
\label{eq:opt_problem}
\max_{\forall \boldsymbol{\theta} \in \Theta} C(\boldsymbol{\theta})  &=1-\frac{\vert\bar{\textbf{$\mathcal{G}$}}_{\boldsymbol{\theta}}\vert}{\vert\textbf{$\mathcal{X}$}\vert}\nonumber\\
{\rm subject \; to}\nonumber\\
 \bar{\textbf{$\mathcal{G}$}}_{\boldsymbol{\theta}} &=f(\textbf{$\mathcal{Y}$}_{\boldsymbol{\theta}},\epsilon)\nonumber\\
\textbf{$\mathcal{Y}$}_{\boldsymbol{\theta}} &=\Phi(\textbf{$\mathcal{X}$},\boldsymbol{\theta})\nonumber\\
 \Theta &=\{\boldsymbol{\theta} \vert \boldsymbol{\theta} \in S,\vert \textbf{$\mathcal{Y}$}_{\boldsymbol{\theta}} \vert \geq \vert \textbf{$\mathcal{X}$}\vert\}\nonumber\\
 S&=\{\boldsymbol{\theta}=(n_1,n_2,...,n_d)\vert n_i \in \mathbb{N}, l \leq n_i \leq u\}
\end{align}
where $\Theta \subset S$ and the sub-space $\Theta$ refers to the feasible domain of the decision space $S$. $f(\textbf{$\mathcal{Y}$}_{\boldsymbol{\theta}},\epsilon)$ generates the factors of $\textbf{$\mathcal{Y}$}_{\boldsymbol{\theta}}$, $\bar{\textbf{$\mathcal{G}$}}_{\boldsymbol{\theta}}$, using the TT-SVD algorithm based on the error bound $\epsilon$. $\Phi(\textbf{$\mathcal{X}$},\boldsymbol{\theta})$ resizes the given tensor \textbf{$\mathcal{X}$} to the shape $\boldsymbol{\theta}$ and enter zero values (dummy elements) if $\vert \textbf{$\mathcal{Y}$}_{\boldsymbol{\theta}} \vert > \vert \textbf{$\mathcal{X}$}\vert$ to fill the rest of the reshaped tensor. The upper limit of the $C(\boldsymbol{\theta})$ is 1. When $0 < C(\boldsymbol{\theta})<1$ the cardinality of the factors are less than that of the data but when $ C(\boldsymbol{\theta})\leq 0$ the memory requirement is inflated and there is no data compression. 

Any shape that results in a tensor $\textbf{$\mathcal{Y}$}_{\boldsymbol{\theta}}$ whose cardinality is smaller than the cardinality of the original given data \textbf{$\mathcal{X}$} ($\vert\textbf{$\mathcal{Y}$}_{\boldsymbol{\theta}}\vert<\vert\textbf{$\mathcal{X}$}\vert$) is infeasible because some data is missed. Furthermore, we allow dummy elements (e.g., zeros) when a possible $\boldsymbol{\theta}$ results in a tensor whose cardinality is greater than that of the original data. Any shape which results in an unnecessary large cardinality is undesirable because it makes the compression less efficient. The objective function defined in Eq.~\eqref{eq:opt_problem} maximizes the compression ratio considering the effect of the added dummy elements. Therefore, the objective function guides the search toward a shape whose cardinality is the closest to that of the data. The definition of the feasible subspace $\Theta$ prevents shapes which have a cardinality smaller than that of the original tensor.

Let us define $E(\boldsymbol{\theta})$ as the relative error measured by the Frobenius norm:
\begin{gather}
E(\boldsymbol{\theta})=\frac{\lVert\textbf{$\mathcal{X}$}-\textbf{$\hat{\mathcal{X}}$}\lVert_F}{\lVert\textbf{$\mathcal{X}$}\lVert_F}, \; {\rm with}\; \textbf{$\hat{\mathcal{X}}$}=\Phi^{-1}(\hat{\textbf{$\mathcal{Y}$}}_{\boldsymbol{\theta}})\; {\rm and}\; 
\textbf{$\hat{\mathcal{Y}}$}_{\boldsymbol{\theta}}=\Psi(\bar{\textbf{$\mathcal{G}$}}_{\boldsymbol{\theta}}),
\end{gather}
where $\Phi(\cdot)^{-1}$ resizes the tensor to the original shape and removes dummy elements if there are any, $\Psi(\cdot)$ generates the approximation tensor $\hat{\textbf{$\mathcal{Y}$}}_{\boldsymbol{\theta}}$ from the factors, and $\bar{\textbf{$\mathcal{G}$}}_{\boldsymbol{\theta}}$ refers to the decomposed factors. 
Since the added dummy elements are zero, then $\lVert\textbf{$\mathcal{X}$}\lVert_F=\lVert\textbf{$\mathcal{Y}$}\lVert_F$ and $\lVert\textbf{$\mathcal{X}$}-\textbf{$\hat{\mathcal{X}}$}\lVert_F \leq \lVert\textbf{$\mathcal{Y}$}-\textbf{$\hat{\mathcal{Y}}$}\lVert_F$. Also, the TT-SVD guarantees that  $\frac{\lVert\textbf{$\mathcal{Y}$}-\textbf{$\hat{\mathcal{Y}}$}\lVert_F}{\lVert\textbf{$\mathcal{Y}$}\lVert_F} \leq \epsilon$. Therefor, if the TT-SVD (described in Algorithm \ref{TT-SVD}) is applied for the decomposition of the reshaped tensor, $E(\boldsymbol{\theta}) \leq \epsilon$  and it is not required to consider the error bound as a constraint in the optimization model.  

\section{Genetic Algorithm for Tensor Shape Search}
We apply a genetic algorithm (GA) to solve the defined optimization model and to find the optimum shape for the tensor decomposition. A pseudo code of the GA for tensor shape search is presented in Algorithm \ref{alg:2} and its key steps are described below.

\subsection{Initialization}The GA starts with generating a set of random shapes (solutions) ${\cal I}=\{\boldsymbol{\theta}_1,\boldsymbol{\theta}_2,...,\boldsymbol{\theta}_m\}$ as an initial population. The initial population is generated by applying a discrete uniform distribution [specifically, $\textbf{unif}(l,u)$] on each variable ($n_i, i=1,2,...,d$) of $\boldsymbol{\theta}_j=(n_1,n_2,...,n_d)$ for $j=1,2,...,m$. Next for each shape $\boldsymbol{\theta}_j$, the TT-SVD is called, and the compression ratio  $C(\boldsymbol{\theta}_j)$ is calculated by Eq.~\eqref{eq:opt_problem}. 

\subsection{Selection}Proportional to the compression ratio, a selection probability is assigned to each shape using the equation below:
\begin{align}
&\Pi(\boldsymbol{\theta}_j)=\frac{C(\boldsymbol{\theta}_j)}{\sum_{j=1}^{m}{C(\boldsymbol{\theta}_j)}}, \; j=1,2,\ldots,m,
\end{align}
where $\Pi(\boldsymbol{\theta}_j)$ is the selection probability of shape $\boldsymbol{\theta}_j$. In the selection process of the GA, $p$ $(p < m)$ shapes are selected as parents. $(p-1)$ solutions are selected based on the probability distribution $\Pi$ (calculated above) with replacement such that the shapes with higher probability ($\Pi$) has more chance to be selected to enter to the parent set. If a solution is selected several times, then several copies of that exist in the parent set. An elitism operation is also applied so that the best shape of the current population (the shape with the maximum compression ratio) is moved to the parent set with probability 1. 

\subsection{Reproduction}During the reproduction process, first the crossover operator is applied. Based on the crossover operator two shapes like $\boldsymbol{\theta}=(n_1,...,n_d)$ and $\boldsymbol{\theta}'=(n'_1,...,n'_d)$ are randomly selected from the parent set and a new trial shape is generated by exchanging the variables of the two selected solution as below:
\begin{equation}
    \boldsymbol{\theta}^{\rm new}=(n_1,...,n_c,n'_{c+1},...,n'_d)
\end{equation}
where $c$ is the crossover point. Next, the mutation operator is applied on the newly generated solution. Based on the mutation operator some of the dimensions (variables) of the newly generated shapes are randomly replaced with applying a discrete uniform distribution $\textbf{unif}(l,u)$. If $\boldsymbol{\theta}=(n_1,...n_i,...,n_d)$ is a newly generated shape by the crossover, the muted shape is $\boldsymbol{\theta}^{{\rm new}}=(n_1,...,n''_i,...,n_d)$ where dimension $i$ is muted. The procedure of selecting parents and generating new solutions continue until $m-p$ new shapes are generated. The compression ratio of the newly generated shapes (new population) are calculated and the selection probabilities are updated. 

\subsection{Iteration and Convergence} The process of selection and reproduction repeats for $T$ iterations. The best final shape is reported as the best (optimum) solution. There is no guarantee that the GA finds an optimum solution but experimental results have shown the effectiveness of the GA in finding a near optimum solution \cite{Bozorg-Haddad}. \cite{Sharapov} presented the stochastic convergence of the elitist GA.
\begin{algorithm}[t]
  \caption{Genetic Algorithm for Tensor Shape Search with Tensor Train Data Compression}
  \label{GA}
  \KwIn{$T$, $m$, $p$}
  \KwOut{Optimum shape $\boldsymbol{\theta}^*$}
  Generate $m$ tentative shapes \\
  \For{$j=1$ to $m$}{
  Run TT-SVD algorithm and Calculate $C(\boldsymbol{\theta}_j)$\\
  }
  $\boldsymbol{\theta}^*$ = the best shape in the current population\\
  \For{$t=1$ to $T$}{
  \For{$j=1$ to $m$}{
  Calculate $\Pi(\boldsymbol{\theta}_j)$\\
  }
  \For{$j=1$ to $p-1$}{
  Select one shape using the distribution $\Pi$\\
  Copy the selected shape to the parent set\\
  }
  Copy the best solution to the parent set\\
  \For{$j=1$ to $m-p$}{
  Generate a new solution using the crossover operator\\
  Mute the newly generated solution using the mutation operator\\
  Run the TT-SVD algorithm for the new shape $\boldsymbol{\theta}_j$ and Calculate $C(\boldsymbol{\theta}_j)$\\ 
  }
  New population = parent set + new solutions\\
  $\boldsymbol{b}$ = the best shape in the current population\\
  \If{$C(\boldsymbol{b})>C(\boldsymbol{\theta}^*)$}{
  $\boldsymbol{\theta}^*=\boldsymbol{b}$\\
  }
  }
  \label{alg:2}
\end{algorithm}


\begin{figure*}
    \begin{minipage}[b]{0.19\textwidth}
    \centering
    \centerline{\includegraphics[width=2.7cm, height=2cm]{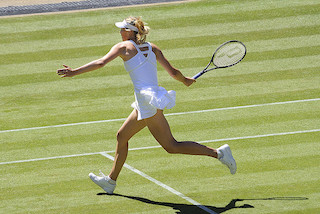}}
    \centerline{Image 1 (214,320,3)}\medskip
    \end{minipage}
    \hfill
    \begin{minipage}[b]{0.19\textwidth}
    \centering
    \centerline{\includegraphics[width=2.7cm, height=2cm]{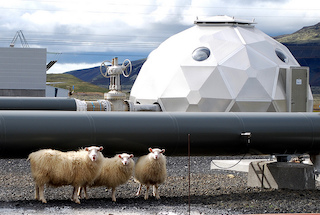}}
    \centerline{Image 2 (215,320,3)}\medskip
    \end{minipage}
    \hfill
    \begin{minipage}[b]{0.19\textwidth}
    \centering
    \centerline{\includegraphics[width=2.7cm, height=2cm]{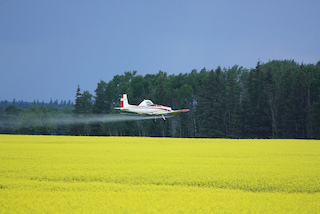}}
    \centerline{Image 3 (214,320,3)}\medskip
    \end{minipage}
    \hfill
    \begin{minipage}[b]{0.2\textwidth}
    \centering
    \centerline{\includegraphics[width=2.7cm, height=2cm]{images/4.jpg}}
    \centerline{Image 4 (213,320,3)}\medskip
    \end{minipage}
    \hfill
    \begin{minipage}[b]{0.19\textwidth}
    \centering
    \centerline{\includegraphics[width=2.7cm, height=2cm]{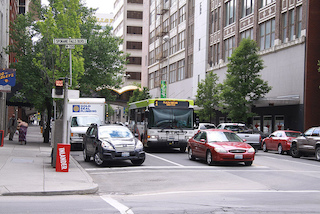}}
    \centerline{Image 5 (214,320,3)}\medskip
    \end{minipage}
    \begin{minipage}[b]{0.19\textwidth}
    \centering
    \centerline{\includegraphics[width=2.7cm, height=2cm]{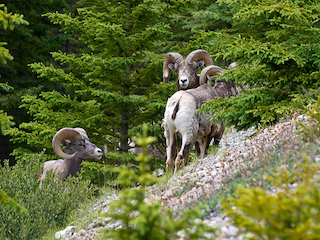}}
    \centerline{Image 6 (240,320,3)}\medskip
    \end{minipage} 
    \hfill
    \begin{minipage}[b]{0.19\textwidth}
    \centering
    \centerline{\includegraphics[width=2.7cm, height=2cm]{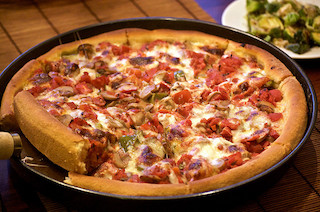}}
    \centerline{Image 7 (212,320,3)}\medskip
    \end{minipage}
    \hfill
    \begin{minipage}[b]{0.19\textwidth}
    \centering
    \centerline{\includegraphics[width=2.7cm, height=2cm]{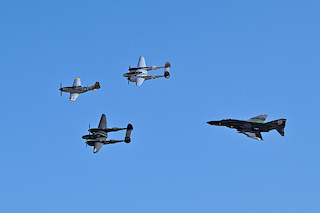}}
    \centerline{Image 8 (213,320,3)}\medskip
    \end{minipage}
    \hfill
    \begin{minipage}[b]{0.19\textwidth}
    \centering
    \centerline{\includegraphics[width=2.7cm, height=2cm]{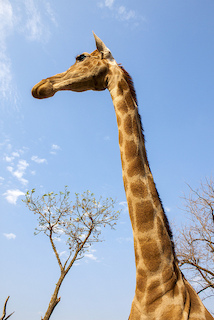}}
    \centerline{Image 9 (320,214,3)}\medskip
    \end{minipage}
    \hfill
    \begin{minipage}[b]{0.19\textwidth}
    \centering
    \centerline{\includegraphics[width=2.7cm, height=2cm]{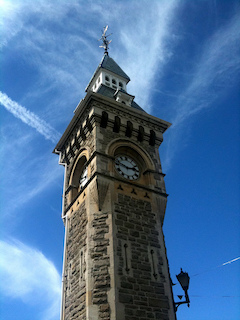}}
    \centerline{Image 10 (320,240,3)}\medskip
    \end{minipage}
    \caption{The arbitrary selected images form COCO data set (the images are not depicted to their correct scale and the numbers written in parenthesis (height, width, depth) refers to the original shape, $\boldsymbol{\theta}$, of the image's data array).}
    \label{fig:1}    
\end{figure*}

\section{Experimental results}

\begin{figure}
    \centering
    \includegraphics[width=\linewidth]{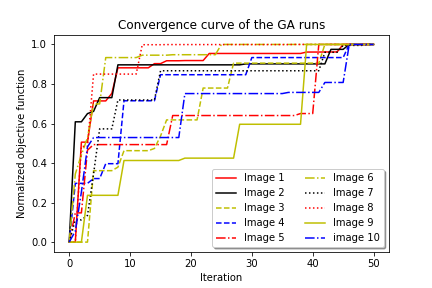}
    \caption{The convergence curve of the GA runs.}
    \label{fig:2}
\end{figure}

We apply the proposed evolutionary tensor shape search linked with the TT-SVD algorithm to decompose some arbitrary RGB images from the Microsoft common objects in context (COCO) data set  \cite{coco} depicted in Fig. \ref{fig:1}. In the experiments, the images are resized such that the longest dimension has 320 pixels with a fixed aspect ratio of the original image. Fig. \ref{fig:1} also shows the original shape (height, width, depth) of the data arrays of the images below them. We compare the decomposition results of the reshaped data with that of the original shapes. In order to have a fair comparison, we set $d=3$ and $l=2$. Therefore, all the optimum shapes are of order three similar to the original shapes. Allowing the GA to change the order may even lead to a better result depending on the data. For each image the GA run for 50 iterations with a population size of 100 and a parent size of 30. Fig. \ref{fig:2} shows the convergence curve of the GA runs for the studied images. 
The result of the experiments for the error bound equal to 0.1 is listed in Table \ref{table1}. In Table \ref{table1}, $\boldsymbol{\theta}^*$ refers to the optimum shape found by the GA. 

\begin{table}
\setlength{\tabcolsep}{0.1 cm}
\caption{The result of the compression of the studied images with their original shape ($\boldsymbol{\theta}$) and the optimum shape ($\boldsymbol{\theta}^*$) for $\epsilon=0.1$}
\label{table1}
\begin{center}
\begin{tabular}{ccc|ccc}
\multicolumn{1}{c}{Image}
&\multicolumn{1}{c}{$C(\boldsymbol{\theta})$\%}
&\multicolumn{1}{c}{$E(\boldsymbol{\theta})$}
&\multicolumn{1}{c}{Optimum shape ($\boldsymbol{\theta}^*$)}
&\multicolumn{1}{c}{$C(\boldsymbol{\theta}^*)$\%}
&\multicolumn{1}{c}{$E(\boldsymbol{\theta}^*)$}
\\ \hline \\
1&72.98&0.0553&(222,16,60)&89.78&0.0694\\
2&75.14&0.0505&(437,8,60)&88.88&0.0701\\
3&94.18&0.0526&(428,10,48)&98.31&0.0680\\
4&82.89&0.0559&(107,16,120)&92.35&0.0696\\
5&62.58&0.0646&(471,8,60)&75.46&0.0702\\
6&17.70&0.0495&(1920,4,30)&58.46&0.0694\\
7&36.07&0.0499&(2270,3,30)&65.71&0.0697\\
8&97.13&0.0583&(71,320,9)&98.65&0.0695\\
9&79.61&0.0505&(193,51,21)&85.61&0.0694\\
10&80.21&0.0504&(349,12,60)&88.52&0.0685\\
\hline \\
\end{tabular}
\end{center}
\end{table}

In Table \ref{table1}, it is seen that for all images the compression ratio of the optimum shape ($\boldsymbol{\theta}^*$) found by the GA is superior to that of the original shape ($\boldsymbol{\theta}$). Also, all the errors are smaller than the error bound $\epsilon$. Although the error slightly increases by improving the compression ratio, the change in the error is negligible and is bounded whereas the improvement in the compression ratio is significant. In Table \ref{table1} it is also seen that the compression ratios of different images vary and it is because the images have different ranks. Regardless of the rank of the images, the proposed method improved the compression ratio of all the studied images. We can conclude that the compression results of the optimum shapes were significantly improved in comparison with that of the original shapes.

\section{Conclusion and future work}
In this work, we have studied the possible effect of the shape of the tensor in data compression using the tensor decomposition. We have formulated the task of finding the optimum shape for the tensor train (TT) decomposition as an optimization model which  maximizes the compression ratio subject to an error bound. We have solved the proposed optimization model using a genetic algorithm (GA) linked with the TT-SVD algorithm. The results have demonstrated the effectiveness of the proposed method in efficiently compressing the given data while keeping the error bounded. 

The study of the effectiveness of the proposed evolutionary tensor shape search for other decomposition methods and improving the efficiency of the optimization algorithm are the subjects of the future studies. 


\vfill\pagebreak


\bibliographystyle{Main}
\bibliography{Main}

\end{document}